\documentclass[fleqn,twoside]{article}
\usepackage[headings]{espcrc2}
\usepackage{graphicx}
\usepackage{rotating}
\usepackage{amsmath,amssymb}
\usepackage{cite,psfrag}
\newcommand{\No}{No.}
\title{Spectra of hadrons and muons in the atmosphere: primary
spectra, characteristics of hadron-air interactions}

\author{A.\,V. Yushkov\address[ASU]{Theoretical Physics Department, Altai State University,\\
Lenin Avenue 61, Barnaul 656049, Russia}%
\thanks{This work was supported by RFBR grants \No\,05--02--27012, \No\,06--02--27113.},
A.\,A. Lagutin\addressmark}

\runtitle{Spectra of hadrons and muons}

\runauthor{A.\,V. Yushkov and A.\,A. Lagutin}

\begin{document}

\begin{abstract}
Self-consistency of interaction models QGSJET~01, SIBYLL~2.1, NEXUS~3.97 and
QGSJET~II is checked in terms of their ability to reproduce simultaneously
experimental data on fluxes of muons and hadrons. From this point of view
SIBYLL~2.1 gives the most acceptable, though not quite satisfactory, results.
Analysis of the situation for muons supports our previous conclusions, that
high-energy muon deficit is due both to underestimation of primary light nuclei
fluxes in direct emulsion chamber experiments and to softness of
$p+A\to\pi^\pm,K^\pm+X$ inclusive spectra in fragmentation region, especially
prominent in case of QGSJET~01 model.
\end{abstract}

\maketitle


\section{Introduction}

At present information on the characteristics of hadronic interactions in
fragmentation region is still scarce or missing and experiments with `roman
pots' are anticipated to improve the situation. Some of this information, in
principle, could be obtained with the use of the data on CR muon and hadron
spectra, provided PCR spectra are known with high precision, but that is not
the case. The obvious obstacle here is that at high energies PCR fluxes
themselves are functionals of various interaction parameters plus their
accuracy is appreciably affected by additional systematic effects. However,
comparison of the hadron and muon fluxes, predicted by different interaction
packages with the experimental data still allows to get information on the
fragmentation particle spectra in the quasi-independent on the PCR fluxes data
way. Besides, as shown below in this paper, our notions on behavior of the PCR
light nuclei spectra can also benefit from such analysis.

\begin{figure}[t]
\psfrag{Energy, GeV}[Tc][Tc][0.8]{$E_\mu$, GeV}
\psfrag{Dif. flux x Energy3.5}[c][c][0.8]{$S_\mu\times E_\mu^{3.5}$, GeV${}^{2.5}$/($\text{cm}^2\cdot\text{s}\cdot\text{sr}$)}
\psfrag{1e2}[r][r][0.65]{$10^2$}
\psfrag{1e3}[r][r][0.65]{$10^3$}
\psfrag{1e4}[r][r][0.65]{$10^4$}

\psfrag{1}[r][r][0.65]{$1$}
\psfrag{2}[r][r][0.65]{$2$}
\psfrag{3}[r][r][0.65]{$3$}
\psfrag{4}[r][r][0.65]{$4$}
\psfrag{5}[r][r][0.65]{$5$}

\includegraphics[width=0.49\textwidth]{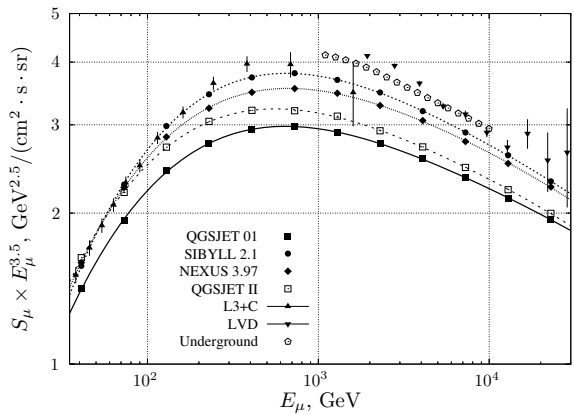}
\caption{Muon spectra at sea level for PCR fits from~\cite{gaisser2002}.
Experimental data: L3+C~\cite{l3_c2004}, LVD~\cite{lvd}, underground~---~the
lower bound of underground measurements~\cite{ya_yaf2006_zemlya_eng}.}
\end{figure}

\section{Muons}
Average numbers of hadrons and muons in EAS were obtained with the help of
one-dimensional hybrid code CONEX~\cite{conex_pylos,conex_2006} in regime of
cascade equations solution for interaction models QGSJET~01~\cite{qgsjet},
SIBYLL~2.1~\cite{sibyll2.1}, NEXUS~3.97~\cite{nexus} and
QGSJET-II-03~\cite{qgsjetii}. In this paper all the results are given for
parameterizations of PCR nuclei spectra from~\cite{gaisser2002} with high
helium flux. Nuclei with $A\geq4$ were treated in the framework of the
superposition model, high accuracy of this approach was confirmed by our
calculations. More details of this check along with complete description of the
calculation procedure will be given elsewhere.

\begin{figure}

\psfrag{energy, gev}[c][c][0.75]{$E_\text{prim}$, GeV}
\psfrag{1e3}[r][r][0.65]{$10^3$}
\psfrag{1e4}[r][r][0.65]{$10^4$}
\psfrag{1e5}[r][r][0.65]{$10^5$}
\psfrag{1e6}[r][r][0.65]{$10^6$}
\psfrag{1e7}[r][r][0.65]{$10^7$}

\psfrag{contribution x 1e11}[c][c][0.75]{Contribution$\times10^{11}/(\text{cm}^2\cdot\text{s}\cdot\text{sr}\cdot\text{GeV})$}

\includegraphics[width=0.49\textwidth]{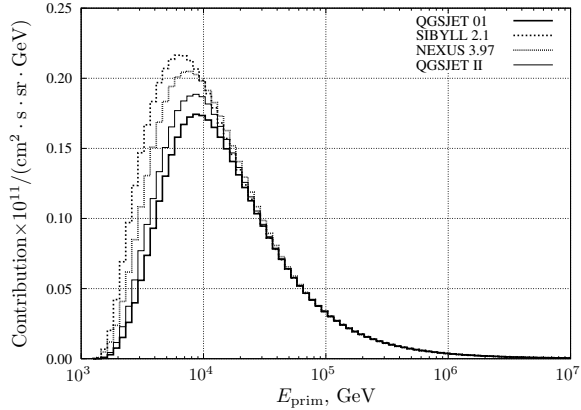}
\caption{Contribution of primary protons with energies $E_\text{prim}$ to the muon
differential spectrum at sea level for $E_\mu=1.29$ TeV.}
\end{figure}

\begin{figure}
\psfrag{energy, gev}[c][c][0.8]{$E$, GeV}
\psfrag{1e3}[r][r][0.65]{$10^3$}
\psfrag{1e4}[r][r][0.65]{$10^4$}
\psfrag{1e2}[r][r][0.65]{$10^2$}
\psfrag{1e0}[r][r][0.65]{$1.0$}
\psfrag{1e-1}[r][r][0.65]{$0.1$}

\psfrag{Edn/dE}[c][c][0.8]{$E\,dn/dE$}
\psfrag{1e-3}[r][r][0.65]{$10^{-3}$}
\psfrag{1e-5}[r][r][0.65]{$10^{-5}$}
\psfrag{1e-4}[r][r][0.65]{$10^{-4}$}
\psfrag{1e-2}[r][r][0.65]{$10^{-2}$}
\psfrag{1e0}[r][r][0.65]{$1$}
\psfrag{1e-1}[r][r][0.65]{$10^{-1}$}
\psfrag{1e1}[r][r][0.65]{$10^1$}

\psfrag{Secondary pions}[c][c][0.65]{$p+A\to \pi^\pm+X$}
\psfrag{Secondary kaons (1/10)}[c][c][0.65]{$p+A\to K^\pm+X\ (\times1/10)$}
\includegraphics[width=0.49\textwidth]{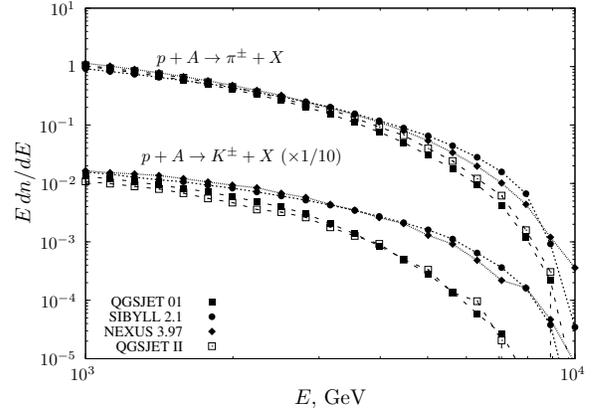}
\caption{Inclusive spectra $p+A\to \pi^\pm+X$ and $p+A\to K^\pm+X$ (scaled down
by 10) for incident proton with energy 10~TeV.}
\end{figure}

Comparison of the calculated muon fluxes with the experimental data, presented
in Figure~1, reveal familiar picture of high energy muon deficit. The reasons
of its appearance were considered in our previous
papers~\cite{ya2004,ya_yaf2006_deficit_eng,ya_pune_p} and they still hold true
regardless of the fact, that three more interaction models were included in our
analysis. All interaction codes, except QGSJET~01, satisfactory describe data
on muon flux only up to $E_\mu\sim$100~GeV and then one by one fail to do it.
Accounting that such muon energies correspond to primary energies above 1~TeV,
studied with balloon(satellite)-borne emulsion chambers, one should simply
relate muon deficit to underestimation of primary light nuclei fluxes, taking
place in these experiments~\cite{ya2004,ya_yaf2006_deficit_eng,ya_pune_p}.
Unfortunately, disagreement between the models in the muon fluxes also appears
at energies around 100~GeV, thus making impossible precise reconstruction of
primary nucleon spectrum for $E_\text{prim}>1$~TeV. In fact, in such conditions
there are no reasons to rule out any of the models, except QGSJET~01, which, as
it was said above, leads to remarkable disagreement with the experiment even in
the range of reliable magnetic spectrometers data on PCR and muon spectra.

To find why the models differ in the predicted muon fluxes let us consider
quite characteristic energy of 1.29~TeV, where discrepancies between the models
reach appreciable values and the data on muons from underground installations
are yet quite reliable. Contributions of primary protons to the differential
flux of muons of the given energy, presented in Figure~2 show, that spread in
muon fluxes between the interaction models is entirely due to uncertainties in
the description of $\pi^\pm,K^\pm$-spectra in fragmentation region
$x=E_{\pi,K}/E_\text{prim}>0.1$ (see Figure~3). Since inclusive muon flux is
sensitive nearly only to the characteristics of the very first primary particle
interaction, hence, the harder these spectra are in the particular model, the
larger muon intensity its use leads to. For the lower values of $x$, i.e. for
$E_\text{prim}>10$~TeV, all the models give practically the same muon yields.
As noted above, in view of uncertain situation with primary spectra for
$E_\text{prim}>1$~TeV, one can not give preference to any of the models in
comparison with the others. If to demand the minimal disagreement with the
direct measurements data on PCR spectra, then obviously SIBYLL~2.1 satisfies
this requirement the best, or, in other words, one may say that it provides the
most acceptable description of $\pi^\pm,K^\pm$ production spectra in $p$-air
collisions in fragmentation region.

\section{Hadrons}

\begin{figure}
\psfrag{energy, gev}[c][c][0.8]{$E_h$, GeV}
\psfrag{flux2.8}[c][c][0.8]{$S_h\times E_h^{2.8}$, GeV$^{1.8}$/($\text{m}^2\cdot\text{s}\cdot\text{sr}$)}
\includegraphics[width=0.49\textwidth]{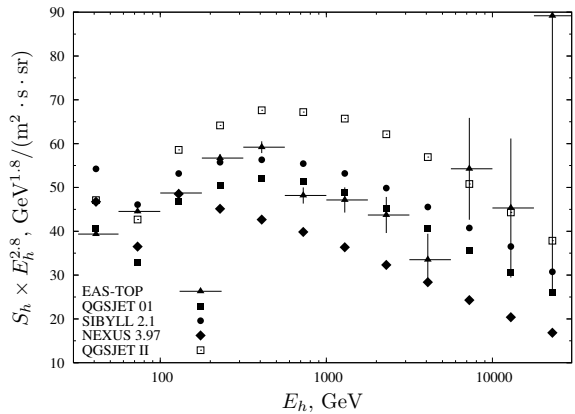}
\caption{Hadron spectra at the EAS-TOP depth $t=820$~g/cm${}^2$.}
\end{figure}

\begin{figure}[t]
\psfrag{energy, gev}[c][c][0.8]{$E$, GeV}
\psfrag{1e3}[r][r][0.65]{$10^3$}
\psfrag{1e4}[r][r][0.65]{$10^4$}
\psfrag{1e2}[r][r][0.65]{$10^2$}

\psfrag{Edn/dE}[c][c][0.8]{$E\,dn/dE$}
\psfrag{1e-3}[r][r][0.65]{$10^{-3}$}
\psfrag{1e-4}[r][r][0.65]{$10^{-4}$}
\psfrag{1e-2}[r][r][0.65]{$10^{-2}$}
\psfrag{1e0}[r][r][0.65]{$1$}
\psfrag{1e-1}[r][r][0.65]{$10^{-1}$}
\psfrag{1e1}[r][r][0.65]{$10^1$}

\psfrag{Secondary pions}[c][c][0.65]{$\pi^\pm+A\to \pi^\pm+X$}
\psfrag{Secondary protons (1/10)}[c][c][0.65]{$p+A\to p+X\ (\times1/10)$}
\psfrag{Secondary neutrons (1/50)}[c][c][0.65]{$p+A\to n+X\ (\times1/50)$}
\includegraphics[width=0.49\textwidth]{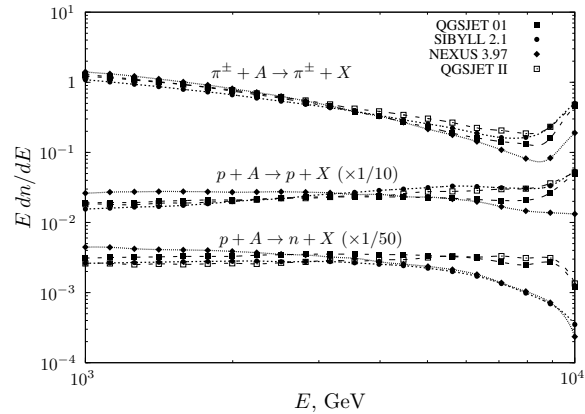}
\caption{Inclusive spectra $\pi^\pm+A\to \pi^\pm+X$, $p+A\to p+X$ (scaled down
by 10), $p+A\to n+X$ (scaled down by 50) for incident particles with energy
10~TeV .}
\end{figure}

Comparison of our calculations with the most recent measurements of inclusive
hadron flux, performed by EAS-TOP group~\cite{eas_top_p}, is presented in
Figure~4. First, let us note the following facts. Below 100~GeV all the
calculated spectra have breaks, because of non-perfect matching of low-energy
model GHEISHA to the high-energy models. Shape of the measured hadron spectra
also breaks at energies above 4~TeV and the data become less definite, thus in
the forthcoming analysis we are going to use data only for energies from
129~GeV to 4~TeV. For these energies QGSJET~01, QGSJET~II and SIBYLL~2.1 quite
reasonably reproduce the shape of the measured hadron spectrum, NEXUS~3.97
leads to spectrum with almost constant power index. One can see, that the most
consistent description of the data for specified energies provide QGSJET~01 and
SIBYLL~2.1. In contrast with the muons there are no energy range, where the
models agree on the hadron fluxes and the reasons of this disagreement are not
as simply to point out as in the case with muons. The most important
characteristics in this analysis are total inelastic cross section, determining
chances of primary particle to survive, shapes of inclusive spectra $p+A\to
p+X$, $p+A\to n+X$, $\pi^\pm+A\to \pi^\pm+X$ in the very forward region,
responsible for substantial process of leading particles production (see
Figure~5 for the listed spectra). Let us briefly outline the major conclusions,
which one may come to in the given situation. NEXUS~3.97 gives the lowest
fluxes as of hadrons in total, so of nucleons and mesons, and this happens in
spite of the lowest inelastic cross-section values. Inclusive spectrum $p+A\to
p+X$ immediately helps to figure out, that incident protons in NEXUS~3.97 have
comparably low chances to save most of their energy in collision and this leads
to such low nucleon flux, the same may be said about meson flux and production
of pions by pions. Similarly, from comparison of the inclusive spectra, it can
be easily understood, why QGSJET~II gives the highest hadron flux. Note, that
SIBYLL~2.1 concedes to QGSJET~II in hadron intensity mostly because of less
effective production of leading neutrons in $p$-air collisions and due to the
larger total interaction cross-section.

Thus, from analysis of the data on hadron flux it is difficult to imply any
strict constraints on inclusive spectra shapes, since mechanism of hadron
spectrum formation is more sophisticated than that in the case of muon
spectrum. SIBYLL~2.1 and QGSJET~01 display quite a different behaviours of the
relevant inclusive spectra and total interaction cross-sections, but both
models almost equally succeed in description of the EAS-TOP data. Alas, even
this conclusion must be taken with care, since it is based on the single set of
data and we have only indirect indications on the accuracy of this set, e.g.
such as agreement of primary proton fluxes, obtained by EAS-TOP and KASCADE
teams (the latter is derived from flux of unaccompanied
hadrons~\cite{kascade_p2004}).

\section{Self-consistency check and conclusions}
Self-consistency implies, that PCR mass composition and spectra, once been 
retrieved from one kind of EAS data with particular interaction model, shall 
bring to satisfactory description of all other types of EAS data with the use 
of this very model. In our case this means, that one can reconstruct flux of 
primary protons, for example, from the data on hadron flux and then to apply it 
as input to get flux of muons. As we have seen, to match the data on muons with 
all the models the primary nucleon flux for $E_\text{prim}>1$~TeV must be 
increased in comparison with the direct measurements data. But, on the 
contrary, to describe EAS-TOP hadron flux with QGSJET~II PCR flux must be 
significantly decreased, to less extent this applies also to SIBYLL~2.1. In the 
case of QGSJET~II this change would result in very large disagreement with the 
data on muons. Since there are almost no need in correction of PCR spectra 
fits~\cite{gaisser2002} for QGSJET~01 to agree with the data on hadron flux, 
this model also fails to satisfy self-consistency conditions. The only model, 
that simultaneously understates fluxes of hadrons and muons is NEXUS~3.97. But, 
even leaving aside misfit of the hadron spectrum shape, any changes in primary 
nucleon intensity for $E_\text{prim}>100$~GeV/n, needed to minimize discrepancy 
with the EAS-TOP data, will immediately lead to disagreement with the data on 
muon fluxes for energies well below 100~GeV. Hence, one may conclude, that 
SIBYLL~2.1 gives the most acceptable overall description of the muon and hadron 
fluxes. Some underestimation of the muon flux for $E_\mu>100$~GeV almost for 
sure should be related to the underestimation of primary nucleon flux for 
$E_\text{prim}>1$~TeV. Overstatement of the hadron flux, which will be 
emphasized by this correction, may be compensated via slight reduction of 
diffractive events fraction.\\ 

Authors greatly acknowledge CONEX team and personally Tanguy Pierog for their
kind permission to use CONEX cross-section tables and for technical support.


\end{document}